\documentclass{aa}                   %  A & A  Latex Macro V8.1
\usepackage{graphicx}                %  Include graphics
\usepackage{txfonts}                 %  Postscript Fonts

\begin{document}

\title{Robotic observations of the most eccentric \break spectroscopic binary in the sky\thanks{Based on
data obtained with the STELLA robotic telescope in Tenerife, an
AIP facility jointly operated by AIP and IAC, as well as on data products from observations made with ESO Telescopes at the La Silla Paranal Observatory under programme IDs 75.C-0733(A) and 60.A-9800(J).}
\fnmsep\thanks{Table \ref{T1} is only available in electronic form
at the CDS via anonymous ftp to cdsarc.u-strasbg.fr (130.79.128.5)
or via http://cdsweb.u-strasbg.fr/cgi-bin/qcat?J/A+A/}
}

\author{K. G.~Strassmeier, M.~Weber \and T.~Granzer}

\offprints{K. G. Strassmeier}

\institute{Leibniz-Institute for Astrophysics Potsdam (AIP), An
der Sternwarte 16, D-14482 Potsdam, Germany\\
\email{[KStrassmeier,MWeber,TGranzer]@aip.de}}

\date{Received ... ; accepted ...}

\abstract{The visual A component of the Gliese~586AB system is a double-lined spectroscopic binary consisting of two cool stars with the exceptional orbital eccentricity of 0.976. Such an extremely eccentric system may be important for our understanding of low-mass binary formation.}{Precise stellar masses, ages, orbital elements, and rotational periods are a prerequisite for comparing stellar observations to angular-momentum evolution models.}{We present a total of 598 high-resolution \'echelle spectra from our robotic facility STELLA from 2006--2012 which we used to compute orbital elements of unprecedented accuracy. New Johnson $VI$ photometry for the two visual components is also presented.}{Our double-lined orbital solution for the A system has average velocity residuals for a measure of unit weight of 41~m\,s$^{-1}$\ for the G9V primary and 258~m\,s$^{-1}$\ for the M0V secondary, better by a factor $\approx$10 than the discovery orbit. The orbit constrains  the eccentricity to 0.97608$\pm$0.00004 and the orbital period to 889.8195$\pm$0.0003\,d. The masses of the two components are 0.87$\pm$0.05~M$_\odot$ and 0.58$\pm$0.03~M$_\odot$ if the inclination is 55$\pm$1.5$^\circ$ as determined from adaptive-optics images, that is good to only 6\%\ due to the error of the inclination although the minimum masses reached a precision of 0.3\%. The flux ratio Aa:Ab in the optical is between 30:1 in Johnson-$B$ and 11:1 in $I$. Radial velocities of the visual B-component (K0-1V) appear constant to within 130~m\,s$^{-1}$\ over six years. Sinusoidal modulations of $T_{\rm eff}$ of Aa with an amplitude of $\approx$55\,K are seen with the orbital period. Component Aa appears warmest at periastron and coolest at apastron, indicating atmospheric changes induced by the high orbital eccentricity. No light variations larger than approximately 4\,mmag are detected for A, while a photometric period of 8.5$\pm$0.2\,d with an amplitude of 7\,mmag is discovered for the active star B, which we interpret to be its rotation period. We estimate an orbital period of $\approx$50,000\,yr for the AB system. The most likely age of the AB system is $\geq$2~Gyr, while the activity of the B component, if it were a single star, would imply 0.5\,Gyr. Both Aa and B are matched with single-star evolutionary tracks of their respective mass. }{}


%\authorrunning{}
%\titlerunning{Robotic observations of the most eccentric spectroscopic binary in the sky}

\maketitle
%------------------------------------------------------------------------

\section{Introduction}

Two investigations independently discovered Gliese\,586A
(=HD~137763, \object{HIP~75718}, G9V + M0V, $P$ = 890\,d)
to be a spectroscopic binary with an extreme orbital eccentricity
of 0.975 (Tokovinin~\cite{tok}, Duquennoy et al.~\cite{duq:may}). This is the highest
eccentricity known among spectroscopic binaries, rivaled by
only a handful of stars, all of which have much longer periods:
HD\,2909 with an eccentricity of 0.949$\pm$0.002 (Mazeh et
al.~\cite{maz:zuc}), HD\,161198 with 0.9360$\pm$0.0007 (Duquennoy
et al. \cite{hd161198}), and possibly also HD\,123949 with an
eccentricity of 0.972$\pm$0.057 (Udry et al. \cite{udry}). The
latter eccentricity is very uncertain due to the long orbital
period of 9200~d and the many data gaps, in particular at the
critical phases. It is therefore probably lower than given, but
certainly remains among the highest known (see the discussion in
Griffin~\cite{g173}). However, Gl586A is outstanding not only
because of its extreme eccentricity, but also because of its
comparably short orbital period and the fact that it is a binary
in a visual system only twenty parsecs away. The visual
B-component is HD\,137778 (\object{HIP\,75722}) separated by 52\arcsec\ in
the sky. It is an early-K star fainter by 0\fm65 mag in the visual and
by 0\fm75 in the blue, according to Duquennoy et
al.~(\cite{duq:may}). The visual AB pair was detected and resolved
in the ROSAT all-sky survey with a luminosity of 42.3\,10$^{27}$
and 46.2\,10$^{27}$~erg\,s$^{-1}$ for A and B, respectively
(H\"unsch et al. \cite{huensch}). Duncan et al. (\cite{duncan})
and Wright et al. (\cite{wright}) determined Ca\,{\sc ii} H\&K
fluxes and found the B-component to be significantly more active
than the A-component, in principal agreement with the X-ray
fluxes. The bona-fide third component, Gl586C, is a
$V$=15\fm4 common-proper-motion star with a mid-M dwarf spectral
type according to its magnitudes (Makarov et al. \cite{mak:zac}).
It is not clear whether it is a true gravitationally bound
component because no sufficient radial velocities exist.

All these circumstances raise serious questions regarding the
evolutionary and dynamical history of this system. As was noted by Duquennoy
et al. (\cite{duq:may}), the separation of the two components at
periastron would be equivalent to that of a circular orbit with a
period of just 3.5\,d. This short timescale is similar to the
expected tidal-shear timescale (Goldman \& Mazeh~\cite{gol:maz})
and suggests tidal interaction at least during periastron passage.
Owing to the orbital period of Gl586A of 890\,d, the two stars
remain well detached, but pass each other at a distance of only
$\approx$10~R$_\odot$ (Duquennoy et al. \cite{duq:may}). If the
same orbital properties also existed during the pre-main-sequence
phase of the two components, the two stars must have been directly
or nearly in contact during periastron passage, or even were too
large to fit into the above separation. Alternatively, the orbit
was not as eccentric in the past as it is today.

A typical cool star in a synchronized binary system with an
orbital period and axial rotation of $\approx$3.5~d would be
expected to show strong chromospheric emission that would be interpreted as
magnetic activity. This is because its convective turn-over time
would be expected to be longer than or of same order as the
rotational period, making internal dynamo action and a surface
magnetic field most likely. However, only very weak Ca\,{\sc ii}
H\& K emission lines, if at all, were seen in a single high-resolution, high
signal-to-noise spectrum taken at orbital phase 0.844 in 1992
(Strassmeier \cite{str}) and no magnetic-field measure exists to
date.

In 2006, we placed Gl586A on the observing schedule of our then
newly inaugurated robotic spectroscopic telescope STELLA in
Tenerife as part of its science-demonstration time. This
paper analyzes data from a total of six consecutive years with a
cadence of roughly one spectrum per week. At an eccentricity of
0.976 and a period of 890~d, the periastron passage of
Gl586A lasts a mere 24 hours. While bad weather
prevented us from seeing the full 2007 event, the Sun was in the line of sight in
2010. The recent passage in February 2012 was only visible for
at most five hours per clear night, but in all allowed for a decent periastron coverage. We include also radial
velocities for Gl586B, the visual companion, and discuss its
variability. Time-series photometry for the two visual components is presented for the first time. The instrument and data are described in
Sect.~\ref{S2}. The new SB2 orbit for component A is presented in
Sect.~\ref{S3} along with new velocities for the visual B
component. We discuss the orbital elements in relation with
the absolute dimensions, ages, and stellar activity of the A and B components in Sect.~\ref{S4}. Sect.~\ref{S5} summarizes our conclusions.

\section{Observations}\label{S2}

\subsection{STELLA/SES spectroscopy in 2006--2012}

\begin{figure}
\includegraphics[angle=0,width=86mm,clip]{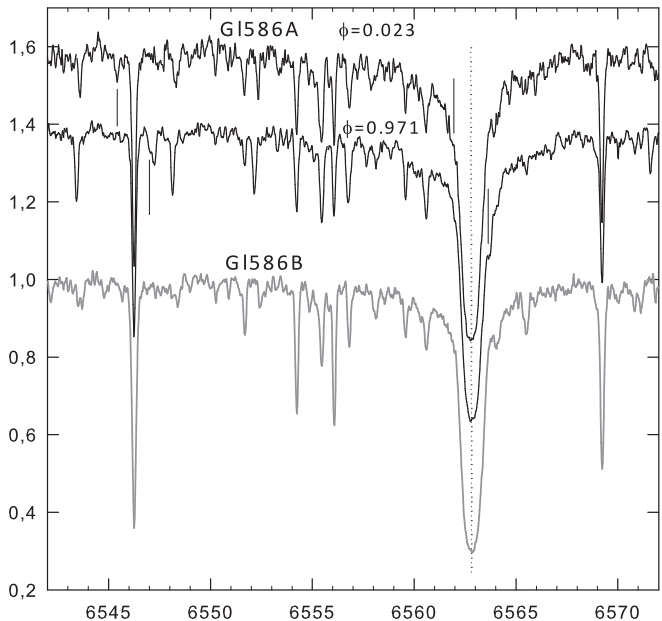}
\caption[ ]{Example spectra of Gl586A and Gl586B for the
wavelength region around Balmer H$\alpha$ . The top two spectra are the
A components shortly before (phase $\phi$=0.971) and after (phase $\phi$=0.023)
periastron. The wavelength position of the weak Ab component is
indicated with dash marks for Fe\,{\sc i}\,6546 and for H$\alpha$ . The bottom spectrum in gray scale is the visual Gl586B component. \label{F1}}
\end{figure}

High-resolution time-series spectroscopy was obtained with the
\emph{STELLA \'Echelle Spectrograph} (SES) at the robotic 1.2-m
STELLA-I telescope in Tenerife, Spain (Strassmeier et
al.~\cite{malaga}, Granzer et al.~\cite{gr:malaga}, Weber et al.~\cite{spie2012}). A total of
598 \'echelle  spectra of Gliese~586A were acquired over the course of almost six
years (2085\,d) between June 26, 2006 (JD\,2,453,913) and
March 12, 2012 (JD\,2,455,998). A total of
39 spectra of Gl586B of similar quality as for the A component were
acquired for this paper. Gl586B was observed between
March 30, 2007 and until May 17, 2012 but with a much lower cadence. The SES is a fiber-fed white-pupil \'echelle
spectrograph with a fixed wavelength format of 388--882\,nm.
Despite increasing inter-order gaps in the red, it continuously records the
range 390--720~nm. Its two-pixel resolution is
$R$=55,000. The CCD was an e2v\,42-40 2048$\times$2048
13.5$\mu$m-pixel device. Figure~\ref{F1} shows spectra of the A stars for the time around periastron and of B for an arbitrary time. Shown is the wavelength range around H$\alpha$\ so that one can recognize the Ab component by eye. Our \'echelle spectra have a useful wavelength coverage of nearly 490\,nm (of which 3\,nm are shown in Fig.~\ref{F1}).

Integrations on Gl586A were set to 2400\,s
and achieved signal-to-noise (S/N) ratios of up to 350:1 per
resolution element, depending on weather conditions. During
periastron passage the exposure time was shortened to 600\,s and had on average S/N ratios of up to $\approx$100:1. Our data are automatically reduced and
extracted using the IRAF-based STELLA data-reduction pipeline (see
Weber et al. \cite{spie}). The two-dimensional data were corrected
for bad pixels and cosmic rays. Bias levels were removed by
subtracting the average overscan from each image followed by the
subtraction of the mean of the (already overscan-subtracted)
master bias frame. The target spectra were flattened by dividing them by the master flat, which had been normalized to unity.
The robot's time series also includes nightly and daily Th-Ar
comparison-lamp exposures for wavelength calibration and
spectrograph focus monitoring. Continuous monitoring of the
environmental parameters inside and outside of the spectrograph
room, most notably temperature and barometric pressure, allows one to
apply proper corrections. For details of the \'echelle data
reduction with particular emphasis on the temperature and pressure
dependencies of the SES, we refer to Weber et al. (\cite{spie}).

Twenty-two radial velocity standard stars were observed with the
same set-up and were analyzed in Strassmeier et al.
(\cite{orbits}). The STELLA system appears to have a zero-point
offset with respect to CORAVEL of +0.503~km\,s$^{-1}$\ (Udry et al. \cite{udry}). The best external
rms radial-velocity precision over the six years of observation
was around 30~m\,s$^{-1}$\ for late-type stars with narrow spectral lines.
All velocities in this paper are on the STELLA zero-point scale if not mentioned otherwise.
The individual velocities are listed in Table~\ref{T1}, available
only in electronic form via CDS Strasbourg.

% --------------------------- Table 1
\begin{table}
\caption{Barycentric radial velocities of Gl586A and
B.} \label{T1}
%\begin{flushleft}
\centering
\begin{tabular}{llllll}
  \hline \noalign{\smallskip}
Instrument & HJD      & $v_{\rm Aa}$ & $\sigma_{\rm Aa}$ & $v_{\rm Ab}$ & $\sigma_{\rm Ab}$ \\
         & & \multicolumn{4}{c}{(km\,s$^{-1}$)}  \\
  \noalign{\smallskip} \hline \noalign{\smallskip}
{\bf A} & & & & &  \\
SES & 2453913.41442 & 5.226 & 0.008 & 10.628 & 0.031 \\
SES & 2453914.40831 & 5.208 & 0.007 & 10.626 & 0.029 \\
SES & 2453915.40807 & 5.191 & 0.008 & 10.719 & 0.031 \\
\dots & \dots & \dots & \dots & \dots \\
{\bf B} & & & & &  \\
SES & 2454189.64040  & 7.898 & 0.003 & & \\
SES & 2454189.67627  & 7.921 & 0.004 & & \\
SES & 2454507.72402  & 8.158 & 0.002 & & \\
\dots & \dots & \dots &&&  \\
\noalign{\smallskip}\hline
\end{tabular}
%\end{flushleft}
%
%{\bf Notes:} Full version available at CDS.
\end{table}

\subsection{Amadeus APT photometry}

Johnson-Cousins $V(I)_C$ photometry of Gl586A and Gl586B was carried out with the {\sl Amadeus} automatic photoelectric telescope (APT) at Fairborn Observatory in southern Arizona from April 6 through May 30, 2012. A total of 1384 measures were made differentially with respect to \object{HD~137666} ($V$=7\fm636, $B-V$=1\fm06, $V-I$=1\fm152) as the comparison star and \object{HD~138425} ($V$=6\fm637, $B-V$=0\fm88, $V-I$=0\fm986) as the check star. Each data point consists of four measurements of the comparison star separated by three measurements of the variable (Gl586A and Gl586B, respectively). At the beginning and end of each block, a check-star measure and a sky-background measure were obtained. A 30\arcsec \ diaphragm was used. Using the standard deviation of the individual target and comparison measures, all data points with an rms of $\geq$0\fm005 were rejected, yielding 678 measures for Gl586A and 706 measures for Gl586B. The standard error of a nightly mean from the overall seasonal mean was 0\fm004 in $V$ and 0\fm006 in $I_C$. For more details we refer to Strassmeier et al. (\cite{apt}) and Granzer et al. (\cite{granzer01}).  From concurrent observations of Johnson standards in $V(RI)_C$, we deduce average values for Gl586A of $V$=6\fm952$\pm$0.004 and $(V-I)_C$= 0\fm91$\pm$0.01 and for Gl586B of $V$=7\fm620$\pm$0.009 and $(V-I)_C$ = 0\fm98$\pm$0.01.

%---------------   F1: example component spectra

\section{New spectroscopic orbital elements}\label{S3}

\subsection{High-precision radial velocities}\label{S31}

The STELLA velocities in this paper were determined from a
simultaneous cross-correlation of 62 \'echelle orders with a
synthetic spectrum. From a pre-computed grid
of synthetic spectra from ATLAS-9 atmospheres (Kurucz~\cite{kur}), we selected
two spectra that matched the respective target spectral
classification. For component Aa, we chose a $T_{\rm
eff}$=5500\,K model, and for Ab a $T_{\rm eff}$=4250\,K
model, both with $\log g$=4.5 and solar metallicities. We
combined the two template spectra to one artificial spectrum using
$v\sin i$ of 2~km\,s$^{-1}$ and a macroturbulence of
3~km\,s$^{-1}$ for both components. We then adjusted the (wavelength-dependent) intensity ratio using the
flux-calibrated spectra from Pickels (\cite{pickels}) multiplied
by 16.5. We then applied a series of radial-velocity
differences between the two components and computed a
two-dimensional cross-correlation function for each of these
shifts. The highest correlation value in the resulting
two-dimensional image corresponds to the measured velocity of
component Aa in one dimension, and the velocity difference of the
two components in the other dimension. To estimate the
uncertainties of this method, we performed a series of 1000 Monte-Carlo
variations of these two-dimensional images using only 63\% of the
original number of \'echelle orders, and found $\sigma$ to be 1/1.349 of
the interquartile range of the resulting distribution.

%---------------   F2: systematics
\begin{figure}
\includegraphics[angle=0,width=86mm, clip]{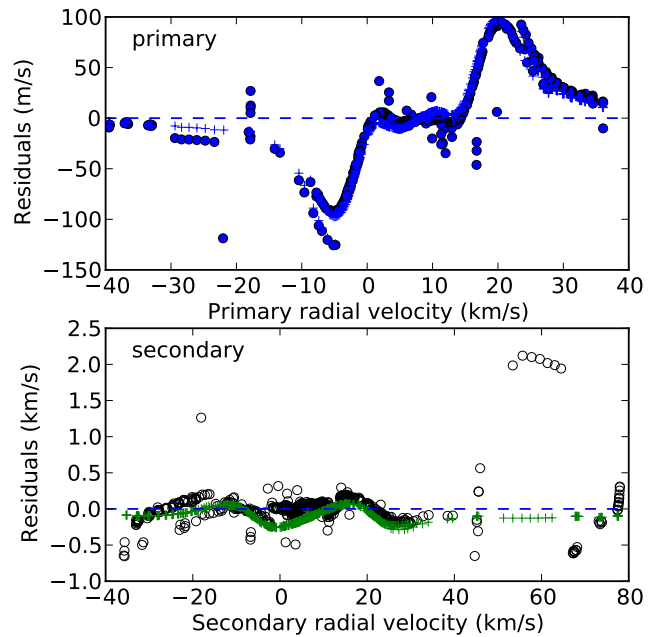}
\caption[ ]{Systematic errors expected from artificial velocities
for Gl586Aa (top panel) and Gl586Ab (bottom panel). A combined synthetic spectrum was
computed for the time of every observed spectrum and radial velocities measured the same
way as for the real observations. Pluses denote the synthetic velocities, circles are for the observed velocities. \label{F2}}
\end{figure}

During the initial year of STELLA operation (2006/7), the external
rms values were significantly higher (120~m\,s$^{-1}$) than
thereafter (30~m\,s$^{-1}$) and also induced a radial velocity zero-point
offset of about 300~m\,s$^{-1}$. In December 2011, we upgraded the
spectrograph's cross disperser but had to use the same optical
camera and CCD until June 2012. During this epoch the CCD recorded
only 45 \'echelle orders instead of the full 82. The wavelength
range was accordingly more narrow, 470--760~nm. 
During these periods, our radial-velocity standard stars exhibit systematic velocity offsets on the order of several 100~m\,s$^{-1}$.
These offsets were applied to the data as well as a barycentric correction.

The last step included corrections for systematic residual
line blending (Fig.~\ref{F2}). We applied a correction similar to
that described in Torres et al. (\cite{torr}) and Weber \&
Strassmeier (\cite{capella}) and then removed the predicted
systematic error from the data. Briefly, we used synthetic
template spectra of the
two binary components with an infinite S/N ratio and rotationally broadened and shifted them to the
expected velocities from an initial SB2 orbit. These spectra have
exactly the same (pixel and phase) sampling as the observations.
They were then flux adjusted and combined into a single spectrum to
mimic an observation of the combined A spectrum. We then processed it
with the same velocity-extraction pipeline as the real
observations and used the differences of the real to the theoretically expected velocities for the corrections. 
These differences for the primary and secondary are compared in the two panels in Fig.~\ref{F2}. 
For Gliese~586A, the residual blending is particularly severe for the secondary velocities because of the high intensity
ratio between components Aa and Ab. It amounts to up to 2~km\,s$^{-1}$.

%---------------   F3: orbit
\begin{figure*}
\includegraphics[angle=0,width=18cm, clip]{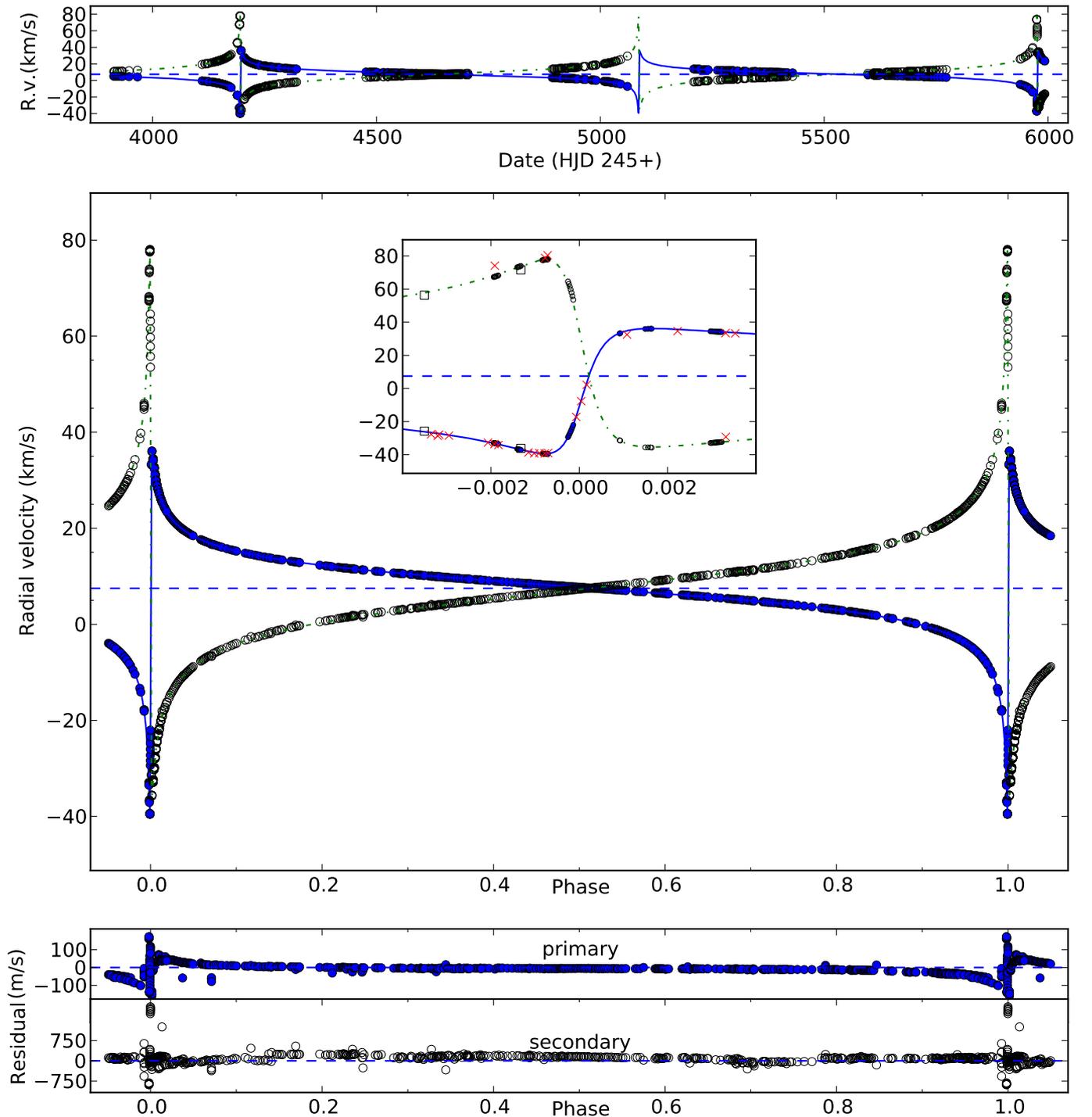}
\caption[ ]{Radial velocities of Gl586A compared with our newly computed orbit. The panels show the
radial velocities versus HJD (top) and orbital phase (middle). The
residuals for both components are shown in the two bottom panels. STELLA data are filled circles for the primary,
star~Aa, and open circles for the secondary, star~Ab. The inset enlarges the time of periastron and additionally  shows two ELODIE points (squares) and some CORAVEL points (crosses). Note that these data were used for the SB1 solution, but not for the SB2 solution. The lines are the elements from Table~\ref{T2}. The horizontal
dashed line is the systemic velocity. \label{F3}}
\end{figure*}

%------------------------------ T2: orbital elements
\begin{table*}[tbh]
\begin{flushleft}
\caption{Spectroscopic orbital elements for Gl586A.}\label{T2}
\begin{tabular}{llll}
\hline \noalign{\smallskip}
Parameter & This paper    & Monte-Carlo & Duquennoy et al.\,(1992)\\
          & w/ rms error  & error       &                         \\
\noalign{\smallskip} \hline \noalign{\smallskip}
$P$ (days)                              & 889.81948$\pm$0.00028$^a$& assumed & 889.62$\pm$0.12\\
$T_{\rm Periastron}$ (HJD)              & 2,454,196.2885$\pm$0.00033 & 0.0019 & 2,447,967.5420$\pm$0.0031\\
$\gamma$ (km~s$^{-1}$)                  & 7.5047$\pm$0.0015 & 0.0039 & 7.323$\pm$0.04$^b$\\
$e$                                     & 0.976081$\pm$0.000012 & 0.00004 & 0.9752$\pm$0.0003\\
$K_{\rm Aa}$ (km~s$^{-1}$)              & 37.844$\pm$0.0058 & 0.028 & 37.14$\pm$0.12 \\
$K_{\rm Ab}$ (km~s$^{-1}$)              & 56.597$\pm$0.031 & 0.034 & 55.50$\pm$0.43 \\
$\omega$ (deg)                          & 255.690$\pm$0.0098 & 0.032 & 253.9$\pm$0.3 \\
$a_{\rm Aa}$~sin~$i$ (10$^6$ km)        & 100.67$\pm$0.029 & 0.11 & 100.95$\pm$0.34\\
$a_{\rm Ab}$~sin~$i$ (10$^6$ km)        & 150.56$\pm$0.090 & 0.15 & 150.87$\pm$1.20 \\
$M_{\rm Aa}$~sin$^3$~$i$ ($M_{\odot}$)  & 0.4782$\pm$0.00068 & 0.0013 & 0.480$\pm$0.030\\
$M_{\rm Ab}$~sin$^3$~$i$ ($M_{\odot}$)  & 0.3197$\pm$0.00033 & 0.00089 & 0.322$\pm$0.018\\
mass ratio, $q\equiv M_{\rm Ab}/M_{\rm Aa}$   & 0.6696$\pm$0.0007 & \dots & 0.670$\pm$0.009 \\
$N_{\rm Aa}$, $N_{\rm Ab}$              & 598, 598 & \dots & 97, 10\\
rms$_{\rm Aa}$ (m\,s$^{-1}$)            & 41 & \dots & 300\\
rms$_{\rm Ab}$ (m\,s$^{-1}$)            & 258 & \dots & 2020\\
\noalign{\smallskip} \hline
\end{tabular}
\end{flushleft}

{\bf Notes:} $^a$from the combined-data SB1 solution as described in the text.\\
$^b$converted to the STELLA zero point for easier comparison.\\
$N$ is the number of radial-velocity measurements used in the
orbit computation.
\end{table*}

\subsection{Double-lined orbit for Gliese~586A}\label{orbit}

We solved for the usual elements of a double-lined spectroscopic
binary using the general least-squares fitting algorithm {\em
MPFIT} (Markwardt \cite{mpfit}). To calculate the eccentric
anomaly, we followed the prescription of Danby \& Burkardt
(\cite{danby:burkardt}).

The initial step is an SB1 solution for component Aa
to constrain the orbital period and verify the high eccentricity.
Not all of the literature CORAVEL data were used, but only the data around the
periastron passage in 1990 (Table~1 in Duquennoy et al.
\cite{duq:may}). These were shifted by --0.503~km\,s$^{-1}$\ to match the
STELLA zero point ($v_{\rm STELLA} - v_{\rm CORAVEL}$=+0.503~km\,s$^{-1}$; 
Strassmeier et al. \cite{orbits}). We also made use of the 17
archival spectra from ELODIE, which we re-analyzed to derive the primary and secondary radial velocities as described above, 
and two from SOPHIE (see Moultaka et~al.~\cite{elodiearchive} for ELODIE \& SOPHIE archives). 
The ELODIE data were concentrated around May 2002.  Two recent SB1 radial
velocities of star Aa from the RAVE survey (Matijevic et al.
\cite{rave}) achieved an rms of around 1--2~km\,s$^{-1}$\ and could not be
used for our high-precision orbit determination. Note
that we also chose not to add the data of Tokovinin (\cite{tok})
because we did not know its zero-point offset with respect to
STELLA or CORAVEL, nor did the authors cover any of the periastron
passages. The combined SB1 solution from STELLA, CORAVEL, ELODIE,
and SOPHIE (total of 695 velocities) converged at an orbital period of 889.81948\,d with an
error of only 0.00028\,d and an rms residual of an observation of
unit weight of 168~m\,s$^{-1}$. Note that we always give orbital periods as
observed and not corrected for the rest frame of the system.
The SB1 solution from STELLA data alone
was very close to the original Duquennoy et al. orbit, but achieved
a significantly better rms of the residuals of 108~m\,s$^{-1}$. Its
eccentricity, 0.97556$\pm$0.00014, was within one $\sigma$ of the
Duquennoy et al. solution, but the longitude of the periastron,
$\omega$, appeared to be offset by 1.5$\degr$ (5~$\sigma$), and the semi-amplitude $K_1$ was larger by 0.51~km\,s$^{-1}$ (4~$\sigma$). 
The mass functions also agreed to within its errors, 0.0521$\pm$0.0002 compared with 0.0523$\pm$0.00056 from STELLA alone.

For the SB2 solution, we use only the STELLA data. First, we corrected the radial velocities of both
components for the difference in gravitational redshift according
to Lindgren \& Dravins (\cite{lindgren:dravins}). For component~Aa
this amounts to 520~m\,s$^{-1}$\ and for component~Ab to 448~m\,s$^{-1}$. A
nominal primary mass and radius were adopted
from Gray~(\cite{gray2005}), while the secondary mass was
iteratively fit to our observed mass ratio in Table~\ref{T2}.
Thus, the net differential velocity offset is 72~m\,s$^{-1}$, and this was
applied to the data.

We kept the orbital period fixed at the value from the
combined-data SB1 solution and then solved for the remaining
elements simultaneously. The final SB2 solution is shown in
Fig.~\ref{F3} and listed in Table~\ref{T2}. Assuming that the fit is of good quality, we derived the element uncertainties by scaling the formal one-sigma errors from the covariance matrix using the measured $\chi^2$ values. For comparison, we also give expected error estimates from a Monte-Carlo reconfiguration of the data.

\subsection{Gliese 586B}

Our STELLA observations of the visual B component are listed in
Table~\ref{T1}. As for all previously published data sets, its
velocity appears quite stable for the time span of the
observation, in our case nearly six years (2007--12). The
data show an rms dispersion of $\approx$130~m\,s$^{-1}$, which is
approximately four times our long-term external precision of a
single measurement. The average velocity is 8.143$\pm$0.013~km\,s$^{-1}$\ (7.640~km\,s$^{-1}$\ in the CORAVEL system).

The basis for the assumption that the Gliese~AB components are indeed
gravitationally bound and not just a common proper-motion pair is
the small velocity difference of 638$\pm$13~m\,s$^{-1}$\ between B and the center of mass of
A. Duquennoy et al. (\cite{duq:may})
provided the first velocities of the B component in March 1978 and
the last in May 1990, a time span of 12.2~yr with an average B
velocity of 7.23$\pm$0.07~km\,s$^{-1}$\ (in their own zero-point system). However, their $\chi^2$ test
suggested a fairly high probability of 0.915 that it is not
constant. Tokovinin (\cite{tok}) just mentioned that the velocity
of Gl586B was constant at 7.6~km\,s$^{-1}$\ (in the Sternberg zero-point system) except for two isolated
measures with 13.4 and 10.6~km\,s$^{-1}$ . No data were given, but the time
must have been between 1986--1991. From this, Tokovinin (\cite{tok}) assumed that
Gl586B could itself be a spectroscopic binary with a high
eccentricity. Earlier, Beavers \& Eitter (\cite{bea:eit}) listed
four Fick measurements from June--July 1982 that had a mean velocity
of 7.7~km\,s$^{-1}$\ with one deviant velocity of 10.5~km\,s$^{-1}$. They also
determined a zero-point shift of $-0.51$~km\,s$^{-1}$\ with respect to
CORAVEL. Nidever et al. (\cite{nid}) listed a mean velocity of
7.752~km\,s$^{-1}$\ (in their zero-point system) with an absolute standard deviation of less than
100~m\,s$^{-1}$\ for a time span of 1479\,d from 1997 to 2001. These
authors also determined the relative zero point of their
iodine-cell measures to CORAVEL to be 53$\pm$7~m\,s$^{-1}$\, which brings their data to within 60\,m\,s$^{-1}$\ of the STELLA data (Table~\ref{T3}). Nordstr\"om
et al. (\cite{nord}) gave a mean velocity of 7.2$\pm$0.3~km\,s$^{-1}$\
from 18 CORAVEL Haute-Provence observations spanning 6014\,d, but
did not list the times. No individual observations were given, but
the authors remarked a fairly low probability of 0.285 that these
velocities were indeed all constant.

Table~\ref{T3} summarizes the mean radial velocities from the literature. The current STELLA data alone do not show significant variations or trends above the rms dispersion of $\approx$130~m\,s$^{-1}$ , in agreement with the data from Nidever et al. (\cite{nid}) from 1997 to 2001. Even though all the mean velocities could indicate a long-term increase of $\approx$0.5~km\,s$^{-1}$\ from 1978 to 2012, the uncertainties of zero-point determinations are such that we believe this to be not conclusive.

%------------------------------ T3: Long-term trend in B velocities
\begin{table}[!tbh]
\begin{flushleft}
\caption{Long-term trend of average Gliese~586B velocities, in
km\,s$^{-1}$ . }\label{T3}
\begin{tabular}{llllll}
\hline \noalign{\smallskip}
Year      & Average     & $N$ & rms   & zero & Reference \\
          & $v_{\rm r}$ \ $^{(a)}$ &     &       & point&           \\
\noalign{\smallskip} \hline \noalign{\smallskip}
1978-79   &7.701    & 9 & 0.26  &+0.503 &  Duquennoy et al.\\
1981-82   &7.238    & 2 & 0.22  &+0.503 &  Duquennoy et al.\\
1982      &7.707    & 4 & \dots &+0.007 &  Beavers \& Eitter \\
1985-87   &7.643    & 2 & 0.06  &+0.503 &  Duquennoy et al.\\
1989-90   &7.940    & 4 & 0.18  &+0.503 &  Duquennoy et al. \\
16yr         &7.703    &18 & 0.3   &+0.503 &  Nordstroem et al.\\
1997-01   &8.202    &12 & 0.10  &+0.450 &  Nidever et al. \\
2007-12   &8.143    &39 & 0.13  &0      &  STELLA, this paper \\
\noalign{\smallskip} \hline
\end{tabular}
\end{flushleft}

{\bf Notes:} $^{a}$In the STELLA zero-point system.\\
$N$ is the number of spectra.
\end{table}

\subsection{Gliese 586C}

Gl586C is listed in the Washington Double Star catalog (Mason et al. \cite{WDS}) as a
possible visual component to the AB pair.  However, it is still not clear whether the bona-fide C component Gl586C = BD-8$^\circ$3984 is physically associated with the AB pair (see
Makarov et al. \cite{mak:zac}). Although its trigonometric
parallax of 47.6~mas (Dahn et al. \cite{dahn}) places it close to
the same distance as the AB pair, it is 20\arcmin\ away in the sky.
The proper motions are slightly uncertain, 63$\pm$30 in right ascension and --347$\pm$30 mas\,yr$^{-1}$ in declination, but formally agree with the A and B components. The available
magnitudes in $VJHK$ suggest a mid-M dwarf ($V$ magnitude of
15\fm4, $B$ magnitude of 17\fm3).

Duquennoy et al. (\cite{duq:may}) cited a single radial velocity of 40.0~km\,s$^{-1}$\
taken in July 1991, from which they concluded that Gl586C is most
likely not physically connected to Gliese~586AB. However, this
needs to be verified. Our own attempt to obtain a
spectrum with SOFIN at the NOT in December 2011 failed due to
mediocre weather and too high an air mass.

\subsection{Gliese 586A orbital inclination}\label{S_incl}

\begin{figure}
\includegraphics[angle=0,width=86mm]{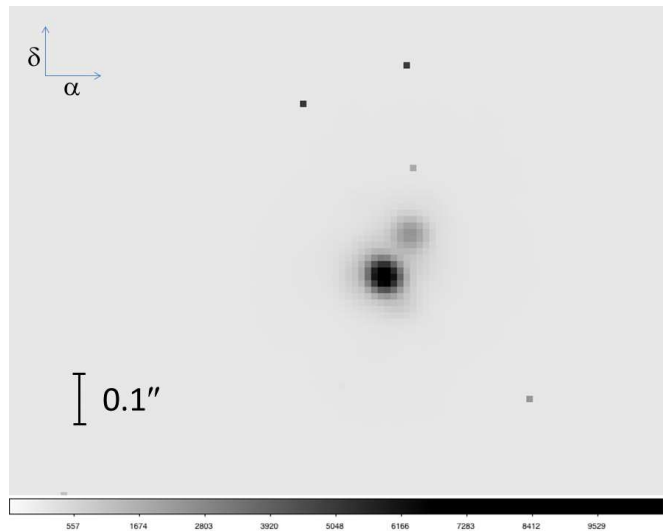}
\caption[ ]{NAOS+CONICA 1.64-$\mu$m image of Gl586A. The Aa component is the
brighter star and Ab (top right) the fainter star. The Aa-Ab separation is
104~mas in this image. \label{F4}}
\end{figure}

Orbit determinations from radial velocity curves leave the inclination of the orbital plane with respect to the line of sight as an unknown. However, separating the Ab component in adaptive-optics images, we were able to indirectly infer the orbital inclination $i$. A series of observations with NAOS+CONICA (NaCo) at the UT-4 VLT is available in the ESO archive (program by Beuzit et al.; see paper by Montagnier et al.~\cite{naos}),
as well as another recent series using the same instrument during technical time. 
Both sets of images show the two A components to be clearly separated by $7.82\pm0.02$~pixel (103.8$\pm$0.3~mas) at a position angle of 328.8$\pm$0.2$^\circ$ in May 2005, and $9.40\pm0.04$~pixel (124.7$\pm$0.5~mas) at a position angle of 336.0$\pm$0.2$^\circ$ in March 2013.
The picture in Fig.~\ref{F4} was taken on May 2, 2005 (2,453,492.3208) with a filter centered at 1.64~$\mu$m and an exposure time of 14~s.

%---------------   F4: NAOS+CONICA

%Using the \emph{Hipparcos} parallax of Gl586A of 48.58$\pm$1.33~mas, the apparent projected A separation is converted to a projected true separation of $D=2.136\pm0.058$~au. Inserting $D$ into $D^2=r^2 (1-\sin^2i \sin^2
Using the \emph{Hipparcos} parallax of Gl586A of 48.58$\pm$1.33~mas, the apparent projected A separation is converted into a projected true separation of $D$. Inserting $D$ into $D^2=r^2 (1-\sin^2i \sin^2
(\omega + \nu))$ after rewriting the equation such that the right-hand side contains only known values,
\begin{equation}
\sin^2i = \left[\sin^2(\omega+\nu)+\left(\frac{D}{a\sin i}\right)^2\right]^{-1} ,
\end{equation}
we find an inclination of $i$=54.3$^\circ$$\pm$3.0$^\circ$\ for the first, and $i$=53.7$^\circ$$\pm$2.7$^\circ$\ for the second set of images, respectively ($r$ being the radius vector between the two stars, $a$ the orbital semi-major axis, $\omega$ the argument of periastron, and $\nu$ the true anomaly). The errors were determined by considering that the orbital parameters are correlated amongst each other.  The error propagation included the covariance according to
\begin{equation}
\sigma^2=\sum_{m,n} \left(\frac{\partial i}{\partial x_m}\right) \left(\frac{\partial i}{\partial x_n}\right){\rm cov}(x_m, x_n) \ .
\end{equation}
Note that this equation reduces to the well-known Gaussian error propagation for uncorrelated values for vanishing non-diagonal elements in the covariance matrix.

Jancart et al.~(\cite{jancart}) detected the motion of the
photocenter of A as seen by {\sl Hipparcos}. Relating this to
the expected center of mass from the spectroscopic orbit by
Duquennoy et al., the authors were able to determine the orbital
inclination to 60.3$^\circ$$\pm$1.8$^\circ$. However, as stated by Jancart et
al.~(\cite{jancart}), their orbital solution failed at least one
consistency test and thus was flagged as uncertain. Therefore, Gl586A did not appear in their main table (appeared only in their Table~4). Because both values, ours from the AO image and Jancart et al.'s from {\sl Hipparcos}, differ by less than $2\sigma$, we are confident of their reality.

Searching the CFHT archive, we found another set of twelve adaptive-optics images. Our analysis of these images led to positions that were more than one order of magnitude less precise than the NaCo images above, and a formal visual orbit fit to these data (keeping the four known parameters from the spectroscopic orbit fixed) did not improve the error bars on the orbital inclination. To estimate the uncertainties introduced by the less precise CFHT values, we calculated the visual orbit with a series of 1000 Monte-Carlo variations of the position measurements, each using only 63\% of the twelve CFHT data points, but adding the two high-quality NaCo data points throughout. Estimating $\sigma$ to be 1/1.349 of the interquartile range of the resulting distribution, we derive $a$=104.2$\pm$2.8\,mas, $i$=55.0$\pm$1.5$^\circ$, and $\Omega$=272.7$\pm$0.9$^\circ$ as
final values for the additional parameters of the visual orbit. We can now determine the distance to be 19.69$\pm$0.65~pc, in good agreement with, but slightly less precise than, the {\sl Hipparcos} value of 20.58$\pm$0.56~pc.

\section{Discussion}\label{S4}

\subsection{Orbit}

The two previous orbit determinations, Duquennoy et al.
(\cite{duq:may}) from CORAVEL observations and Tokovinin
(\cite{tok}) from Sternberg observations, had rms
residuals for the primary of 0.30 and 0.40~km\,s$^{-1}$, respectively,
but 2.02~km\,s$^{-1}$\ for the secondary (only Duquennoy et al.
\cite{duq:may}). Tokovinin (\cite{tok}) did not see traces of the
secondary and presented an SB1 solution.

The precision of an individual STELLA measurement is superior to
any of the two previous data sets, owing mostly to spectrograph
stability and higher spectral resolution. Considering the expected
systematic errors due to the velocity measuring technique, and a
relative gravitational redshift between the components, our rms
residual from the orbital solution is just 41~m\,s$^{-1}$\ for the primary and
258~m\,s$^{-1}$\ for the secondary. If we exclude the data points during periastron $\pm0\fp05$, which have higher than normal rms due to the rapid velocity change, then the rms is 13~m\,s$^{-1}$\ for the primary and 103~m\,s$^{-1}$\ for the secondary.  Formally, this is better by
approximately a factor 23 and 20 for the primary and the
secondary, respectively, than the CORAVEL orbit.
Sampling done by a robot is expectedly unprecedented and a total
of 598 velocities for both components were
available. However, because our total time coverage
is shorter than what was available for the initial CORAVEL orbit,
12~yr versus 6~yr, we chose to include the CORAVEL data from around periastron passage in 1990
and a few ELODIE and SOPHIE measures from 2002 for the period determination. This gave us a baseline of 18~yr or approximately 7.4 orbital revolutions.

The semi-amplitude of component~Aa from our orbit is larger
by 704~m\,s$^{-1}$\ (6~$\sigma$) than the Duquennoy et al. (\cite{duq:may}) value, but more precise by a
factor of 10, while for component~Ab it is larger by 1097~m\,s$^{-1}$\ (2.5~$\sigma$) and
more precise by a similar factor. The same can be said for
the eccentricity, which is confined to within 3.8\,10$^{-5}$ at
$e=0.97608$ and more precise by a factor 10
than the initial CORAVEL orbit. Because of these very low rms
numbers and the fact that an inspection of the radial-velocity
curves in Fig.~\ref{F3} shows residual systematic errors only at
the level of around the accuracy of a single measurement, we conclude
that our orbit is also more accurate by similar factors.

The STELLA $\gamma$ velocity of the A pair from 2006--2012 is higher by 182~m\,s$^{-1}$\ than the
Duquennoy et al. value of 6.82$\pm$0.04~km\,s$^{-1}$\ from 1978--1990 (i.e. 7.323~km\,s$^{-1}$\ in
the STELLA system). The mean velocity of the B
component is also higher, on average by $\approx$500~m\,s$^{-1}$,
instead of being decreased by a similar amount, as would be expected for a
physical pair. If both velocity offsets are real, this would be puzzling unless B itself is an undetected
long-period spectroscopic binary with a full radial-velocity amplitude of smaller or around the rms of our data or that of Nidever et al. (\cite{nid}) ($\approx$100~m\,s$^{-1}$). Alternatively, the A and B stars are gravitationally unbound and are just a co-moving pair of similar stars.

In pursuing the question whether the A--B system is truly bound or just a co-moving pair, we converted the projected  A-B separation into 1066$\pm$35\,au, which equals a minimum separation between A and B, and compared it to the observed radial velocities. The independently measured \emph{Hipparcos} distances for the A and B stars are consistent with each other within their errors (Table~\ref{T4}). We can safely assume that they are equally far away, and together with the observed radial velocity difference of B with respect to the center-of-mass velocity of A, that is 638$\pm$13~m\,s$^{-1}$,  would formally allow for a bound orbit. For $e=0$ and the mass estimates given in Table~\ref{T4}, the unprojected orbital velocity of the B component with respect to A would indeed equal 640~m\,s$^{-1}$, which is very close to the observed relative radial velocity of 638\,m\,s$^{-1}$. The expected period for such a circular orbit would still be $\approx$50,000\,yr, beyond hope of being observable within the few years of coverage. We can independently estimate the tangential velocity of the AB system from the differences of their proper motions to be between $+0.49\pm0.3$~km\,s$^{-1}$\ and $-1.75\pm1.2$~km\,s$^{-1}$. The total velocity turns out to be almost exactly the escape velocity of the AB system. Although the errors are too large to give conclusively constrained velocities, one can calculate the maximum separation between A and B for a just marginally bound system. This, in turn, can be converted into a minimum orbital A-B inclination of $78^\circ$. If $i=90^\circ$, the eccentricity of the A-B orbit would be required to be as high as $e=0.98$, similar to that of the A system.

The bona-fide C component, if at all part of the Gl586 system, has a minimum separation to A of 25,500\,au. For a circular orbit this transforms to an exorbitant long orbital period of $\geq27$~Myr.

%---------------   F5: PARSES output
\begin{figure}
\includegraphics[angle=0,width=86mm]{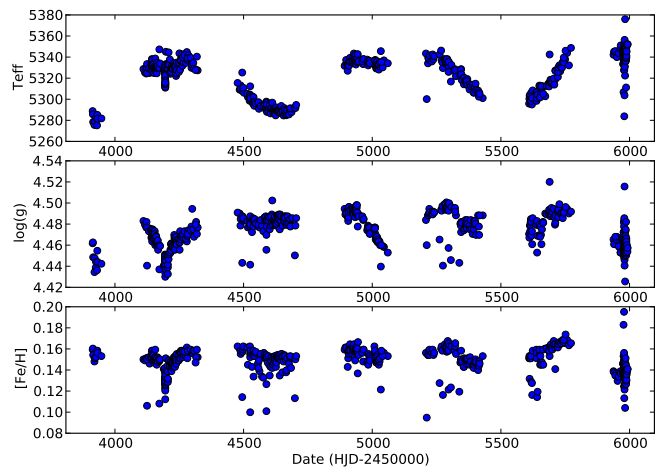}
\caption[ ]{Atmospheric parameters of Gl586Aa from the PARSES analysis. Note a $\approx$55~K amplitude of the effective temperature with the period of the orbit in the top panel.
\label{F5}}
\end{figure}

\subsection{Physical properties}

%-------------------------   Table 4:  Astrophysical values of Gliese 586AB
\begin{table}
\caption{Summary of astrophysical properties of Gliese~586AB.
\label{T4}}
\begin{flushleft}
 \begin{tabular}{llll}
  \hline\noalign{\smallskip}
  Parameter                & Aa & Ab & B \\
  \noalign{\smallskip}\hline\noalign{\smallskip}
  $V$, mag		& 7.01 & 10.15 & 7.62 \\
  Spectral class           & G9V & M0V & K0-1V \\
  Parallax, mas            & 48.58$\pm$1.33 & 48.58 & 48.80$\pm$0.89 \\
  P.m.$\alpha$, mas/yr     & 87$\pm$2    & \dots & 82$\pm$1\\
  P.m.$\delta$, mas/yr     & --374$\pm$1 & \dots & --356$\pm$1\\
  $T_{\rm eff}$, K         & 5330$\pm$70 & (4000) & 5110$\pm$50 \\
  $\log g$, cm\,s$^{-2}$   & 4.48$\pm$0.03 & \dots& 4.34$\pm$0.03 \\
  $[$Fe/H$]$, solar        & +0.15$\pm$0.01 & \dots & +0.24$\pm$0.02\\
  $v\sin i$, km\,s$^{-1}$          & 1.2$\pm$0.5 & 3$\pm$1 & 1.5$\pm$0.5 \\
  Inclination, $^\circ$    & 55$\pm$1.5 & \dots & 12-27  \\
  Rotation period, d       & n.d. & \dots & 8.5$\pm$0.2 \\
  Radius, R$_\odot$        & 0.92$\pm$0.04 & 0.59$\pm$0.03  & 0.78$\pm$0.04 \\
  Luminosity, L$_\odot$    & 0.61$\pm$0.04 & 0.080$\pm$0.006 & 0.37$\pm$0.02 \\
  Mass, M$_\odot$          & 0.87$\pm$0.05 & 0.58$\pm$0.03 & 0.85$\pm$0.05 \\
  $\log$ Li abundance      & n.d. & n.d. & n.d. \\
  Age, Gyr                 & $\geq$2 & \dots & (0.5)\\
   \noalign{\smallskip}\hline
 \end{tabular}
\end{flushleft}

n.d.: not detected.
\end{table}

We applied our tool PARSES (``PARameters from SES''; Allende-Prieto~\cite{all04}) to all individual STELLA spectra. PARSES is implemented as a suite of Fortran programs within the STELLA data analysis pipeline and is based on the synthetic-spectrum-fitting procedure described in Allende-Prieto~(\cite{all04}). Synthetic spectra are computed and pre-tabulated for relative logarithmic metallicities of --2.5 to +1.0, logarithmic gravities between 0 to 5.0, and temperatures between 3000\,K to 7000\,K for a wavelength range of 380--920\,nm. All calculations were based on MARCS model atmospheres (Gustafsson et al. \cite{marcs}) with the VALD3 line list (Kupka et al. \cite{vald3}; with updates on some specific $\log gf$ values) and were fixed with a microturbulence of 1.1~km\,s$^{-1}$\ for Gl586 A and B. Macroturbulence was set to 3~km\,s$^{-1}$. This grid was then used to fit 50 selected \'echelle orders of each STELLA/SES spectrum. The four parameters $T_{\rm eff}$, $\log g$, $[$Fe/H$]$, and $v\sin i$ were solved for simultaneously in all orders. Internal errors of the fits were estimated using the original noise in the spectra. We verified this approach by applying it to the ELODIE library (Prugniel \& Soubiran \cite{elodie}), and used linear regressions to the offsets with respect to the literature values to correct our PARSES results.

\begin{figure}
\includegraphics[angle=0,width=86mm, clip]{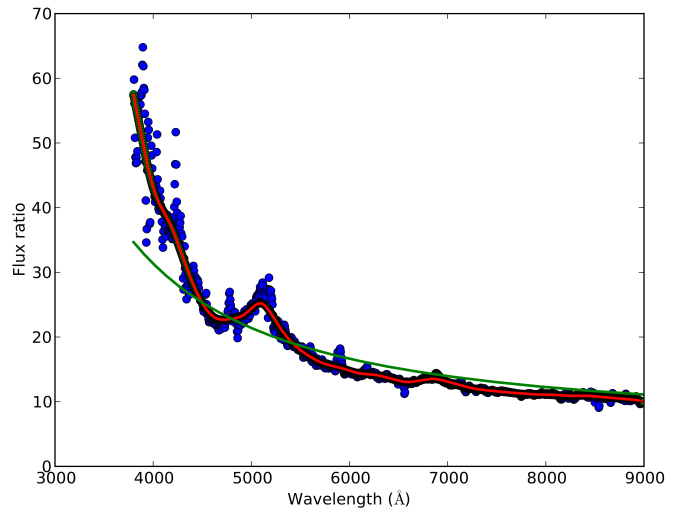}
\caption[ ]{Flux ratio of components Aa/Ab versus wavelength. The dots are the observed ratios binned to 5\,\AA\  and the thick line is its functional. The thin line is the ratio of two black bodies of 5250\,K and 4000\,K with  radii appropriate for the two components.}\label{F6}
\end{figure}

PARSES treats the Gl586A spectrum as a single-star spectrum and per default does not extract the secondary. Values would be formally ``combined values'', but nevertheless are quite precise for the primary due to the faintness of the secondary and the fact that for most of the spectra the two components are unblended.  Our next step in the analysis was the removal of the secondary spectrum from all (598) Gl586A spectra before the PARSES analysis. This was done by using the same synthetic spectrum for Ab as for the systematic-error correction of the radial velocities in Sect.~\ref{S31} and with the same wavelength-dependent scaling as given in Fig.~\ref{F6}. 

The PARSES results for Aa are shown in Fig.~\ref{F5} as a function of time. Systematic changes are obvious and a Lomb-Scargle periodogram of the $T_{\rm eff}$ results reveals a clear period of 885$\pm$35\,d with a full amplitude of $\approx$55\,K and an rms of below 10\,K (correlation coefficient of 0.89), which is equal to the orbital period within its errors. The highest temperature occurs repeatedly at periastron and the lowest temperature at apastron. The same is noted for the $\log g$ and $[$Fe/H$]$ time series, but with significantly smaller relative amplitudes; 0.05$\pm$0.03(rms) in $\log g$ and 0.03$\pm$0.02(rms) in $[$Fe/H$]$. Table~\ref{T4} lists the long-term average values with their grand rms errors.

%---------------   F6: flux ratio vs. wavelength

We used the NAOS+CONICA images to perform psf-modeling and aperture photometry of the residuals for  both components for two sets of six and ten exposures at 1.26~$\mu$m and 1.64~$\mu$m, respectively. The averaged magnitude differences Gliese\,586 Aa-minus-Ab are 1\fm890$\pm$0.013 at 1.26~$\mu$m and 1\fm480$\pm$0.015~mag at 1.64~$\mu$m. For comparison, the flux ratio primary to secondary at optical wavelengths from the dip of the radial-velocity cross-correlation function is 13.5$\pm$0.5 in the $R$ band and 18$\pm$1 in the $V$ band, that is, a magnitude difference of approximately 2\fm8 in $R$ and 3\fm15 in $V$.

If we adopt the orbital inclination of 55$\pm$1.5$^\circ$, the individual masses for Aa and Ab are 0.87$\pm$0.05~M$_\odot$ and 0.58$\pm$0.03~M$_\odot$, respectively. These are masses with an accuracy of only 6\%, but are by far dominated by the error of the inclination. Note that the minimum masses are good to $\approx$0.3\%. The primary mass fits an $\approx$G7-8 dwarf star with a nominal radius of 0.9~R$_\odot$, but even better a K0V from its effective temperature of 5250\,K according to the table in Gray~(\cite{gray}). The MILES atmospheric parameter library (Cennaro et al. \cite{miles}) suggests a G8-9V classification, but with a slightly smaller $\log g$ of +4.32. We adopted G9V for the remainder of the paper and note that this value is not based on a direct spectral classification. The secondary mass suggests an $\approx$K7-M0 star with a predicted radius of 0.57~R$_\odot$ according to the models of a 4~Gyr, $[$Fe/H$]$=0 isochrone from Baraffe et al. (\cite{bar}) as given in Fernandez et al. (\cite{fer:lat}).  With a mass-radius relation of $R\propto M^{0.79}$, we would expect radii of 0.92~R$_\odot$ and 0.67~R$_\odot$ for the two components, respectively. A more direct determination of the radii comes from the combined magnitudes of the two Gl586A stars and the known distance. Together with their flux ratio of 18$\pm$1 at $V$-band wavelengths, the individual apparent magnitudes translate into absolute magnitudes of $M_V$=5\fm44$\pm$0\fm06 for Aa and 8\fm58$\pm$0\fm07 for Ab. For normal main-sequence stars this implies G7-8V and M0V according to the tables in Gray~(\cite{gray}). Similarly, Gl586B has $M_V$=6\fm06$\pm$0\fm05 and thus approximately fits a canonical K0-1V star. With the bolometric corrections from Flower (\cite{f96}) based on the $T_{\rm eff}$ from our spectrum synthesis and a solar bolometric magnitude of 4\fm72, the three stars' bolometric magnitudes correspond to luminosities of 0.610$\pm$0.041~L$_\odot$, 0.080$\pm$0.006~L$_\odot$, and 0.37$\pm$0.02~L$_\odot$ (Aa, Ab, B). These luminosities suggest radii of 0.92$\pm$0.04~R$_\odot$, 0.59$\pm$0.03~R$_\odot$, and 0.78$\pm$0.035~R$_\odot$ for Aa, Ab, and B, respectively, again with the effective temperatures from Table~\ref{T4}. Errors are simply propagated from $L$ and $T_{\rm eff}$. The M0 secondary is indeed slightly larger than predicted from the Baraffe et al. tracks for solar metallicity, in agreement with recent radii measurements of eclipsing late-K and M dwarfs (see L\'opez-Morales \cite{lop}, Fernandez et al. \cite{fer:lat}).
%However, assuming the same metallicity of $[$Fe/H$]$=--0.10 for the M0 star as for the G8 star brings this radius %closer to

Koen et al. (\cite{koen}) listed spectral types of G9V for component A (i.e., Aa and Ab combined) and K2V for component B from their $UBVRIJHK$ photometry, both in overall agreement with our spectroscopic values. The Aa and the B star differ in their effective temperatures by only 200~K and agree in their (logarithmic) metallicities and their gravities to within 2\,$\sigma$ of the measurement error, but their luminosities differ significantly.

\subsection{Stellar rotation and magnetic activity}

The projected rotational line broadening is determined to be $v\sin i$=1.2$\pm$0.5~km\,s$^{-1}$\ for Aa, 3$\pm$1~km\,s$^{-1}$\ for Ab, and 1.5$\pm$0.5~km\,s$^{-1}$\ for B with adopted macroturbulences of 2~km\,s$^{-1}$\ for all three stars following the recipe in Fekel~(\cite{fcf}). Microturbulent broadening was assumed to be 1.0~km\,s$^{-1}$. The PARSES spectrum fit yielded average $v\sin i$'s of 2.0$\pm$0.4\,km\,s$^{-1}$\ for Aa and 2.4$\pm$0.4\,km\,s$^{-1}$\ for B, in reasonable agreement with Fekel's recipe. The high orbital eccentricity suggests a rather speedy pseudo-synchronous rotation period for Aa of $\approx$3.4~d according to the tidal-friction theory of Hut (\cite{hut}), which would be just 2.6\%\ of the orbital angular velocity. This clearly disagrees with the measured $v\sin i$, which suggests a rotation
period closer to $\approx$30~d if $i=55\degr$. Most likely, the Aa star is a strongly asynchronous rotator.  Note that Wright et al. (\cite{wright}) had determined \emph{expected} periods from the strength of the Ca\,{\sc ii} S-index for Aa and B based on the rotation-activity relation from Noyes et al. (\cite{noyes}) of 39~d and 9~d, respectively.  While they had observed the B component 15 times, the A component(s) had only a single H\&K measurement (A on JD2,450,277; B between 2,450,277--835).

We analyzed our APT photometry with the Lomb-Scargle periodogram (Lomb~\cite{lomb}; Scargle~\cite{sca}) and also the minimum string-length algorithm following Dworetsky~(\cite{MSL}). The $V$-band rms scatter of our data sets were 4~mmag for the combined A light curve, but 9~mmag for the B component, while it was in the $I$-band 10~mmag for A and 7~mmag for B. The larger-than-normal $V$-band rms scatter of Gl586B indicates that it is photospherically mildly active, in agreement with the Ca\,{\sc ii}~H\&K observations of Duncan et al. (\cite{duncan}) and Wright et al. (\cite{wright}), and our own spectra. The 4-mmag rms scatter for A agrees with the expected observational error if the star is constant. Its Lomb-Scargle periodogram shows a period of $\approx$1.2\,d (in addition to the dominating $1$\,d aliasing from the window function). Its full amplitude of $\approx$0\fm003 is below the intrinsic scatter of the data and this period is consequently judged spurious. No other variability could be detected.

%---------------   F7: APT Lichtkurven
\begin{figure}
\includegraphics[angle=0,width=86mm, clip]{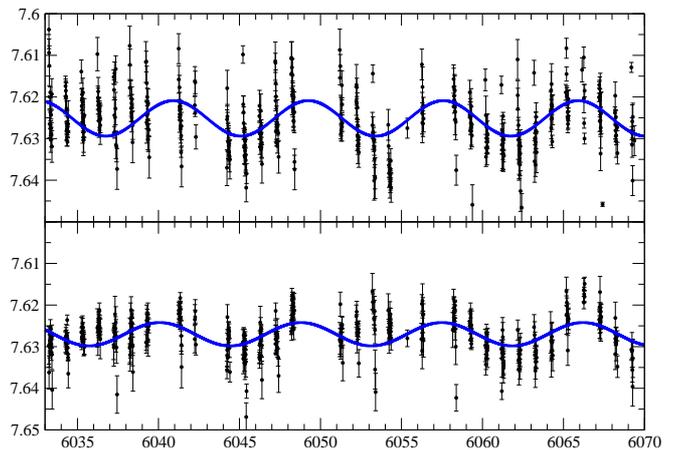}
\caption[ ]{APT $V$-band photometry of Gl586B. The top panel shows magnitudes with respect to the check star HD\,138425, the lower panel with respect to the comparison star HD\,137666. The time axis is given in truncated Julian date 245+. The harmonic fits with periods of 8.33\,d (top panel) and 8.71\, d (bottom panel) are overplotted.}\label{F7}
\end{figure}

The Gl586B-minus-comparison data after JD\,2,456,032 show a periodicity of 8.71\,d with a full amplitude of 0\fm0056. Given the external uncertainty of 0\fm0049 from the check-minus-comparison light curve, this period does not appear to be significant. However, the differential data with respect to the check star show a similar period of 8.33\,d with an amplitude of 0\fm0085 at an intrinsic scatter of 0\fm007. The respective two light curves are shown in Fig.~\ref{F7}. Expected (minimum) period errors are determined from refitting of synthetic photometry (see also Strassmeier et al.~\cite{orbits}). This method estimates confidence intervals by synthesizing a large number of data subsets out from the original data by adding Gaussian random values to the measurements proportional to the actual rms of the data. By generating $10^5$ synthetic data sets and calculating a Lomb-Scargle periodogram on all of them, we adapted the standard deviation of its periods as the expected error. Errors retrieved in this way were 0.09\,d for the variable-minus-comparison data and 0.1\,d for the variable-minus-check data. However, these errors are lower limits because the method does not take into account systematic noise. We conclude that the two periods are not significantly different and thus adopted their arithmetic average as the best value from the current data, that is $P_{\rm rot}$(Gl586B)=8.5$\pm$0.2\,d. Note that this period agrees well with the expected period of 9\,d from the rotation-activity relation.

The range of inclinations of the rotation axis of B would be just 12--27$^\circ$ from the measured $v\sin i$ range and the 0.78-R$_\odot$ radius from the Boltzmann-relation and above $P_{\rm rot}$ (with the error of $v\sin i$ dominating over the error of $R/P_{\rm rot}$ by a factor 10). All inclination values within this range are equally likely, but do not overlap with the 55$\pm$1.5$^\circ$ inclination of the A orbital plane. Because Gl586B shows moderately strong H\&K emission, it is actually unusual that the photometric light amplitude is so small, $\approx$7\,mmag, and barely detectable from the ground. The low inclination of the rotation axis would be a natural explanation for this.

\subsection{Ages}

\begin{figure}
\includegraphics[angle=0,width=86mm,clip]{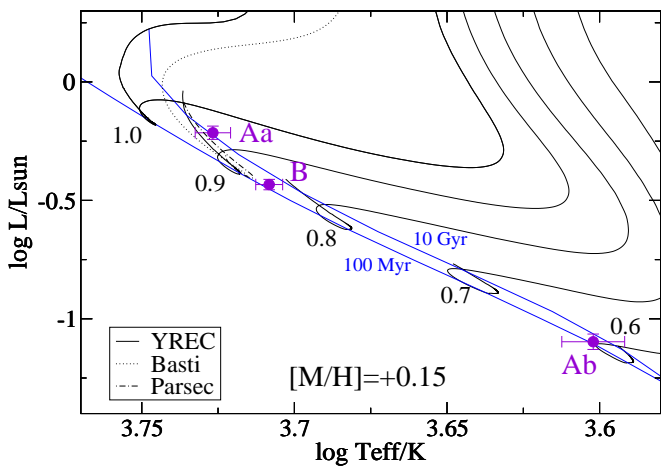}
\caption[ ]{Comparison of the Gliese\,586 stars (dots marked Aa, Ab, and B) with theoretical evolutionary tracks and isochrones for a metallicity of 0.15 matching Gl586Aa. The full lines are taken from the grid of Spada et al. (\cite{spada}) computed with YREC. Shown are pre-main sequence and post-main sequence tracks for the masses labeled and two isochrones of age 0.1 and 10\,Gyr. For comparison, the dotted and dash-dotted lines are the 0.9-M$_\odot$ main-sequence tracks from {\sl Basti} (Pietrinferni et al. \cite{piet}) and {\sl Parsec} (Bressan et al. \cite{parsec}), respectively. }\label{F8}
\end{figure}

Figure~\ref{F8} compares the position of the three Gliese-586 stars with theoretical evolutionary tracks and isochrones in the $L - T_{\rm eff}$ plane. Tracks and isochrones are taken from Spada et al. (\cite{spada})
and were computed for [Fe/H]=$+0.15$ with an updated version of the Yale Rotating stellar Evolution Code (YREC; Demarque et al. \cite{yrec}) in its non-rotating configuration. These models also include moments of inertia of the radiative and convective zones and convective turnover timescales, in view of their application to studies of the rotational and magnetic evolution of low-mass stars. The most important difference to the previous YREC version is the treatment of the equation of state and the use of outer boundary conditions based on updated non-gray atmospheric models.

The locations of Aa and B in the H-R diagram match their corresponding main-sequence YREC tracks within their errors. The same holds true for both the {\sl Basti} (Pietrinferni et al. \cite{piet}) and the {\sl Parsec} tracks (Bressan et al. \cite{parsec}). 
%Given the orbitally derived mass for the Aa star of 0.90$\pm$0.09\,M$_\odot$
%The Ab-star mass (0.6\,M$_\odot$), on the other hand, would need to be 200\,K cooler to match the theoretical main-sequence track within its errors. 
The B-component is an effectively single star without a mass determination. It is approximately matched with a track of mass 0.85$\pm$0.05~M$_\odot$ for the metallicity of $0.15$. 
%Unfortunately, isochrones are generally of only limited use for low-mass stars like the Gliese\,586 components and the YREC isochrones would only support a pre-main-sequence status for Aa and B. In this respect, we are forced to conclude that binary evolution in the past must have significantly affected the AB stars.

%---------------   F8: HRD

No lithium absorption line at 670.78\,nm is detected for Gl586A or for B down to our detection limit of 2--3~m\AA . This effectively rules out a pre-main-sequence nature of the two stars and points to an older age than the 0.6\,Gyr of the Hyades (Sestito \& Randich \cite{ses:ran}). The
three $^{12}$C$^{14}$N lines around 800.35~nm are clearly detectable with a combined
equivalent width of 61~m\AA\ for Aa and 77~m\AA\ for B. However, the equivalent width of
the $^{13}$C$^{14}$N line at 800.46~nm is not detected in either of the two stars, and we may again state just a lower limit of $\approx$2--3~m\AA . Thus, the $^{12}$C/$^{13}$C ratio must be lower than $\approx$24 for Aa and lower than $\approx$30 for B, which is significantly lower than the presumed initial solar ratio of about 90 (see Lambert \& Ries \cite{lam:rie}) and indicative of evolved main-sequence stars. If we assume that all three Gliese-586 stars fit in with other local dwarfs within 15~pc of the Sun, defined by the S$^4$N survey (Allende Prieto et al.~\cite{all}), we might expect an age younger than that of the Sun, though. The age concentration of the S$^4$N survey peaks sharply at $\approx$1~Gyr, with the Sun being among the oldest stars in our galactic neighborhood. However, the formal range of ages is 0.16--10~Gyr for the whole sample. In any case, the two moderately high $^{12}$C/$^{13}$C ratios together with the statistical results from S$^4$N suggest a ``few Gyr'' age for Gl586Aa and B.

An independent cross-check is provided by the gyrochronological approach of Barnes (\cite{barnes}). Although it is not applicable to tidally interacting binaries, it is interesting to state what the effective temperatures and rotation periods would predict. Because Gl586Aa has no detected rotation period, we rely on its indirectly determined value based on the measured equatorial rotational velocity and the stellar radius (based on the orbital inclination and additionally assuming that the spin axis is perpendicular to the orbital plane). With the values in Table~\ref{T4} the most likely rotation period for Aa is 30~d, while we measured a period of 8.5\,d for B. Then, the Barnes (\cite{barnes}) relation would suggest gyroages of approximately 5~Gyr for Aa but 0.5~Gyr for B, which would agree with the observed magnetic activity levels of both stars, but would obviously disagree with the common assumption that the AB components are coeval. Ages of 2~Gyr for Aa and 0.5~Gyr are obtained when the red-dwarf calibration of Engle \& Guinan (\cite{eng:gui}) is applied. Most likely, this just demonstrates that gyrochronology is not applicable to binaries. Even for a single star with planets, gyroages are probably biased due to expected star-planet interactions, as demonstrated recently for HD\,189733 (Santapaga et al. \cite{san:gui}).

\section{Conclusions}\label{S5}

We confirmed the extreme eccentricity of Gliese~586A and presented a new and much more precise and accurate spectroscopic orbit. Currently there exists no formation scenario for a binary system with an orbital period as short as 889.8195$\pm$0.0003\,d and an eccentricity as high as 0.97608$\pm$0.00004 in a triple system. An evolutionary scenario based on a theory for the coupling of the envelope shear with a constant turbulent viscosity (Zahn~\cite{zahn}) would require an initial semi-major axis of about 1~pc and an even higher eccentricity. Such a large separation between the components would make the survival of the system very unlikely. Goldman \& Mazeh~(\cite{gol:maz}) favored a quadratic reduction of the convective-envelope viscosity over time that solves this problem and results in initial conditions very similar to the present one. Of course, it does not explain how such a system is formed in the first place and whether, for instance, a capture mechanism is a viable option or can be excluded. A crucial piece in this puzzle is the existence of the close-by and very similar Gliese~586B star at the same distance and location and with a similar proper motion. 
%In such a scenario it is actually not surprising that the two bright components, Aa and B, do not match single-star evolutionary tracks. 
Our new data for Gliese~586A and B agrees with the assumption that B is gravitationally bound to the A stars, but does not conclusively exclude the common proper-motion scenario. The fact that the $\gamma$-velocity of the A system and the velocity of the visual B star are both increasing over time would agree if B itself were a long-period very eccentric spectroscopic binary just like A. We now have evidence that the inclination of the rotation axis of B is very low, $i\approx12^\circ\dots27^\circ$. If perpendicular to the orbital plane of a hypothetical (unseen) secondary star, radial-velocity variations would be very difficult to measure and thus not contradict the current measurements of B.

Observations of the faint, bona-fide C component are still needed to confirm its membership in the system.

\acknowledgements{STELLA was made possible by funding through the
State of Brandenburg (MWFK) and the German Federal Ministry of
Education and Research (BMBF). The facility is a collaboration of
the AIP in Brandenburg with the IAC in Tenerife. We thank Federico Spada for making the YREC tracks available to us prior to publication. We also thank Manfred Woche and Emil Popow and his team as well as
Ignacio del Rosario and Miquel Serra-Ricart from the IAC Tenerife
day-time crew. We also thank Ilya Ilyin for his attempt to observe
Gliese~586C at the Nordic Optical Telescope. Furthermore we thank Thierry Forveille to point out to us the observations in the CFHT archive, and the referee for his constructive comments.
This research used the facilities of the Canadian Astronomy Data Centre operated by the National Research Council of Canada with the support of the Canadian Space Agency.
}

%-------------------------------------------------------------------------
\end{document}